\documentclass[preprint,showpacs,preprintnumbers,amsmath,amssymb]{revtex4}

\usepackage{graphicx}
\usepackage{dcolumn}
\usepackage{bm}

\renewcommand{\bar}[1]{\overline{#1}}

\begin{document}

\title{Photon-meson transition form factors of
light pseudoscalar mesons}
\author{Bo-Wen Xiao}
\affiliation{Department of Physics, Peking University, Beijing
100871, China\\
Department of Physics, Columbia University, New York, 10027, USA}
\author{Bo-Qiang Ma}
\altaffiliation{Corresponding author}\email{mabq@phy.pku.edu.cn}
\affiliation{ CCAST (World Laboratory), P.O.~Box 8730, Beijing 100080, China\\
Department of Physics, Peking University, Beijing 100871, China}

\begin{abstract}
The photon-meson transition form factors of light pseudoscalar
mesons $\pi ^{0}$, $\eta $, and $\eta ^{\prime }$ are
systematically calculated in a light-cone framework, which is
applicable as a light-cone quark model at low $Q^{2}$ and is also
physically in accordance with the light-cone pQCD approach at
large $Q^{2}$. The calculated results agree with the available
experimental data at high energy scale. We also predict the low
$Q^{2}$ behaviors of the photon-meson transition form factors of
$\pi ^{0}$, $\eta $ and $\eta ^{\prime }$, which are measurable in
$e+A(\mbox{Nucleus})\rightarrow e+A+M$ process via Primakoff
effect at JLab and DESY.
\end{abstract}





\vfill



\vfill


\vfill



\vfill 



\maketitle

\section{Introduction}

The meson-photon and photon-meson transition form factors contain
interesting physics concerning the QCD structure of both photons
and mesons. The pion-photon transition form factor provides a very
simple example for the perturbative QCD (pQCD) analysis to
exclusive processes, and was first analyzed by Brodsky and Lepage
\cite{Lep80} at large $Q^{2}$. It has been shown \cite{Cao96} that
the applicability of pQCD can be extended to lower $Q^{2}$ around
a few $\mbox{GeV}^{2}$ by taking into the transverse momentum
contributions in both hard scattering amplitude and pion wave
function. In our recent study \cite{Xiao03} within light-cone
quark model, it is shown that the pion-photon transition form
factor is identical to the photon-pion transition form factor when
taking into account only QCD and QED contributions. Therefore the
formalism that applies to the pion-photon transition form factor
is also applicable to the photon-pion transition form factor.
Taking the minimal quark-antiquark Fock states of both the photon
and pion as their wave functions, we could calculate the
photon-pion transition form factor by using the Drell-Yan-West
assignment. This framework is applicable at low $Q^{2}$ as a
light-cone quark model approach, and it is also physically in
accordance with the light-cone pQCD approach at large $Q^{2}$.
Thus we can describe the photon-pion form factors at both low
$Q^{2}$ and high $Q^{2}$ within a same framework. The purpose if
this work is to apply this framework \cite{Xiao03} for a
systematic description of the photon-meson transition form factors
of pseudoscalar mesons $\pi ^{0}$, $\eta $, and $\eta ^{\prime }$,
at both $Q^{2}\rightarrow 0$ and $Q^{2}\rightarrow \infty $
limits, and to make predictions in a wide $Q^{2}$ range.

The photon-meson transition form factor $\gamma ^{\ast }\gamma
\rightarrow M$ can be realized in $e+e\rightarrow e+e+M$ or
$e+A(\mbox{Nucleus})\rightarrow e+A+M$ processes. The $\gamma
^{\ast }\gamma \rightarrow M$ transition form factors of $\pi
^{0}$, $\eta $, and $\eta ^{\prime }$ at medium to high $Q^{2}$
have been measured at Cornell~\cite{CLEO98} and at
DESY~\cite{Beh91} through the $e^{+}+e^{-}\rightarrow
e^{+}+e^{-}+M$ process, while the latter process
$e+A(\mbox{Nucleus})\rightarrow e+A+M$ is convenient to provide
measurement of the photon-meson transition form factors at low
$Q^{2}$. Moreover, high precision measurements of the
electromagnetic properties of these pseudoscalar mesons via
Primakoff effect are proposed by PrimEx Collaboration at the
Thomas Jefferson National Accelerator Facility (JLab) \cite{Gan},
which would give the experimental value of transition form factors
$F_{\gamma ^{\ast }\gamma \rightarrow M}(Q^{2})$ of $\pi ^{0}$,
$\eta $, and $\eta ^{\prime }$ at low
$Q^{2}~(0.001-0.5~\mbox{GeV}^{2})$, and lead to a clarification on
the obvious disagreement between the former Primakoff experiment
and collider cases in the measurements of $\Gamma (\eta
\rightarrow \gamma \gamma )$ and a more precise determination of
the $\eta\mbox{-}\eta ^{\prime }$ mixing angle. Similar
measurements can be also performed by HERMES Collaboration at
Deutsche Elektronen-Synchrotron (DESY) \cite{hem98}. Therefore,
theoretical predictions at low $Q^2$ are necessary and essential
for comparison with future experimental measurements.

It is well known that the physical $\eta $ and $\eta ^{\prime }$
states dominantly consist of a flavor $SU(3)$ octet $\eta _{8}$
and singlet $\eta _{0}$ in the $SU(3)$ quark model, respectively.
The usual mixing scheme reads:
\begin{equation}
\left(
\begin{array}{c}
\eta \\
\eta ^{\prime }
\end{array}
\right) =\left(
\begin{array}{cc}
\cos \theta & -\sin \theta \\
\sin \theta & \cos \theta%
\end{array}
\right) \left(
\begin{array}{c}
\eta _{8} \\
\eta _{0}
\end{array}
\right).
\end{equation}
Using different sets of experimental data, we recalculate the value
of the mixing angle $\theta $ by employing the limiting method
developed by Cao-Signal \cite{Cao99}. Our results are also
compatible with other approaches \cite{Feld99} for the mixing angle
and scheme.

In general, people use the chiral perturbation theory \cite{Chpt}
or some other methods \cite{nChpt} which deal with current quark
masses in order to take the chiral symmetry and chiral anomaly
into account, since the chiral symmetry predominates the
$\pi^{0}$($\eta$, $\eta^{'}$)$\gamma\gamma$ vertex at large $Q^2$
\cite{Cheng}, and chiral anomaly determines the $\pi^{0}$($\eta$,
$\eta^{\prime }$) transition form factors at $Q^2 =0$
(Eqs.~(\ref{pi_gamma}-\ref{eta_gamma})). In addition, the chiral
perturbation theory is also very useful and effective in
discussing the $\eta $ and $\eta ^{\prime }$ mixing properties
\cite{Mixing}. Since we are consistently using the valence quark
masses in the light-cone treatment to the form factor calculation,
it is not very applicable to start with current quark mass within
the chiral symmetry and investigate the chiral limits in the
transition form factor computation. However, our main purpose of
this paper is to employ the new light-cone $\gamma\rightarrow
q\overline{q}, s\overline{s}$ wave functions~\cite{Qiao00,Xiao03,
Ani04} to compute the transition form factors of the light mesons.
Moreover, we considered the chiral symmetry when we choose $\eta $
and $\eta ^{\prime }$ mixing scheme, and took the chiral limit
approximation when we try to determine and fix the parameters.
Therefore our results respect the chiral symmetry and its breaking
at some extent. Phenomenologically, we could give the predictions
of the $\eta $ and $\eta^{\prime }$ mixing angle within the
light-cone formalism, as well as the photon-meson transition form
factor which is applicable at both low and high energy scales.

This paper is organized as follows. In section 2 we present the
formalism for the photon-meson transition form factor using the
minimal quark-antiquark Fock states of the photon and pion as wave
functions. In section 3, we will introduce the $\eta\mbox{-}\eta
^{\prime }$ mixing scheme used in our calculation. In section 4,
we calculate systematically the photon-meson transition form
factors of $\pi^0$, $\eta$, and $\eta'$, and show that the
calculated results agree with the available experimental data at
medium to large $Q^2$ scale. We also predict the low $Q^{2}$
behaviors of the photon-meson transition form factors of $\pi
^{0}$, $\eta $, and $\eta ^{\prime }$, which are measurable in
$e+A(\mbox{Nucleus})\rightarrow e+A+M$ process via Primakoff
effect at JLab and DESY. In section 5, we present a brief summary.

\section{Formalism of photon-meson transition form factor}

\begin{figure}
\par
\begin{center}
\scalebox{0.5}[0.5]{\includegraphics{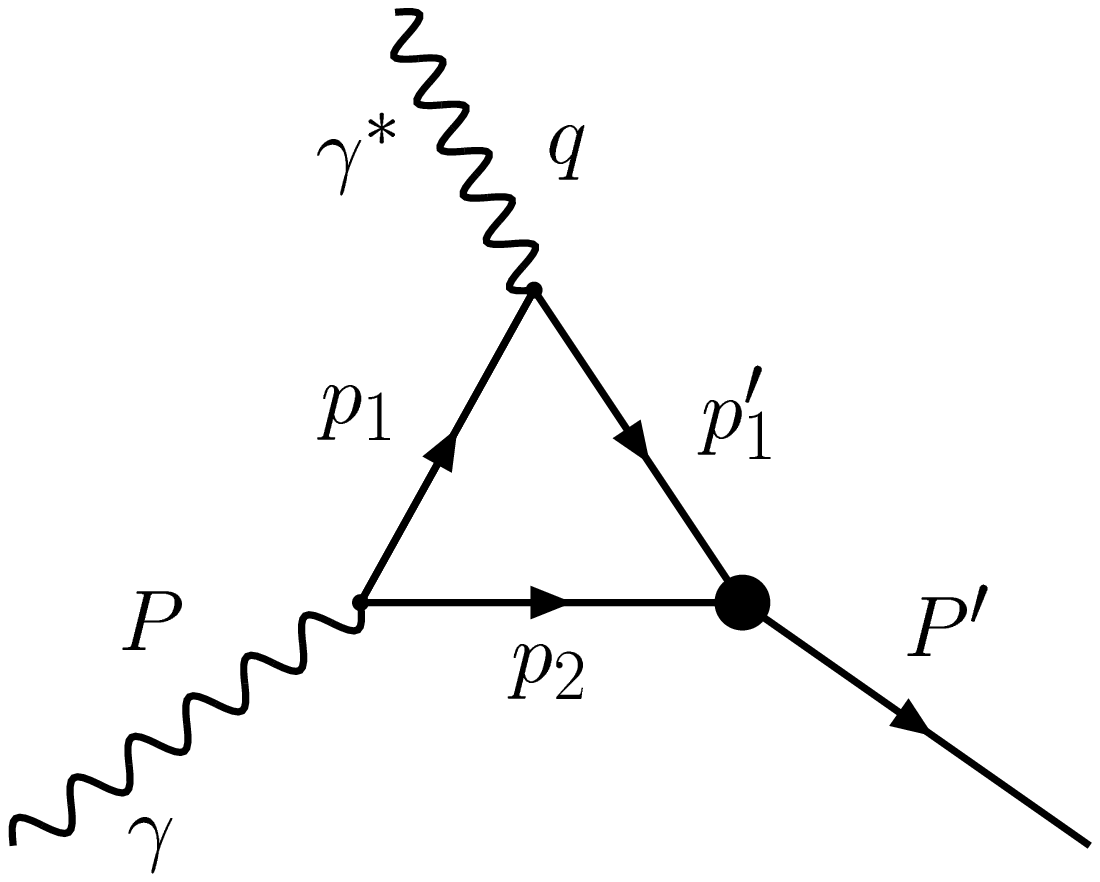}}
\end{center}
\caption[*]{\baselineskip13pt The diagram for the contribution to the
transition form factor $F_{\protect\gamma ^{\ast }\protect\gamma \rightarrow
M}$. The arrows indicate the particle moving directions.}
\label{tranfig}
\end{figure}

We work in the light-cone formalism \cite{Bro89}, which provides a
convenient framework for the relativistic description of hadrons in
terms of quark and gluon degrees of freedom, and for the application
of perturbative QCD to exclusive processes. The transition form
factor $F_{\gamma ^{\ast }\gamma \rightarrow M}$ ($M=\pi ^{0}$,
$\eta $, and $\eta ^{\prime }$), in which an on-shell photon is
struck by one off-shell photon and decays into a meson, as
schematically shown in Fig.~\ref{tranfig}, is defined by  the
$\gamma ^{\ast }\gamma M$ vertex,
\begin{equation}
\Gamma _{\mu }=-ie^{2}F_{\gamma ^{\ast }\gamma \rightarrow
M}(Q^{2})\varepsilon _{\mu \nu \rho \sigma }p_{M}^{\nu }\epsilon ^{\rho
}q^{\sigma },
\end{equation}
in which $q$ is the momentum of the off-shell photon,
$Q^{2}=-q^{2}=\mathbf{q}_{\perp }^{2}-q^{+}q^{-}=\mathbf{q}_{\perp
}^{2}$ is the squared four momentum transfer of the virtual
photon, and $\epsilon $ is the polarization vector of the on-shell
photon. We choose the light-cone frame
\begin{equation}
\left\{
\begin{array}{lll}
P & = & (P^{+},\frac{q^{2}+\mathbf{q}_{\perp }^{2}}{P^{+}},\mathbf{0}_{\perp
}), \\
P^{\prime } & = & (P^{\prime +},\frac{M^{2}}{P^{\prime +}},\mathbf{q}_{\perp
}),\\
q & = & (0,\frac{Q^{2}}{P^{+}},\mathbf{q}_{\perp }),
\\
p_{1} & = & (xP^{+},\frac{\mathbf{k}_{\perp }^{2}+m^{2}}{xP^{+}},\mathbf{k}%
_{\perp }) \\
p_{2} & = & ((1-x)P^{+},\frac{\mathbf{k}_{\perp }^{2}+m^{2}}{(1-x)P^{+}},-%
\mathbf{k}_{\perp }), \\
p_{1}^{\prime } & = & (xP^{\prime +},\frac{\mathbf{k}_{\perp
}^{\prime 2}+m^{2}}{xP^{+}},\mathbf{k^{\prime }}_{\perp }).
\end{array}%
\right.
\end{equation}

Instead of calculating the diagram directly, we introduce the
quark-antiquark wave function of the photon \cite{Xiao03} by
calculating the matrix elements of
\begin{equation}
\frac{\overline{u}(p_{1}^{+},p_{1}^{-},\mathbf{p}_{1\perp
})}{\sqrt{p_{1}^{+} }}\gamma \cdot \epsilon
\frac{v(p_{2}^{+},p_{2}^{-},\mathbf{p}_{2\perp })}{
\sqrt{p_{2}^{+}}},
\end{equation}%
which are the numerators of the wave functions corresponding to
each constituent spin $S^{z}$ configuration. The two boson
polarization vectors in light-cone gauge are $\epsilon ^{\mu
}=(\epsilon ^{+}=0,\epsilon ^{-}, \mathbf{\epsilon }_{\perp })$,
where $\mathbf{\epsilon }_{\perp \uparrow ,\downarrow }=\mp
\frac{1}{\sqrt{2}}(\widehat{\mathbf{x}}\pm \widehat{
\mathbf{y}})$. To satisfy the Lorentz condition $k_{photon}\cdot
\epsilon =0, $ the polarizations have the relation $\epsilon
^{-}=\frac{2\mathbf{\ \epsilon }_{\perp }\cdot \mathbf{k}_{\perp
}}{k^{+}}$ with $k_{hoton}$, thus we have
\begin{equation}
\left\{
\begin{array}{lll}
\Psi _{R}^{\uparrow }(x,\mathbf{k}_{\perp },\uparrow ,\downarrow
)=-\frac{ \sqrt{2}(k_{1}+ik_{2})}{1-x}\varphi _{\gamma }, & \left[
l^{z}=+1\right] &
\\
\Psi _{R}^{\uparrow }(x,\mathbf{k}_{\perp },\downarrow ,\uparrow
)=+\frac{
\sqrt{2}(k_{1}+ik_{2})}{x}\varphi _{\gamma }, & \left[ l^{z}=+1\right] &  \\
\Psi _{R}^{\uparrow }(x,\mathbf{k}_{\perp },\uparrow ,\uparrow
)=-\frac{
\sqrt{2}m}{x(1-x)}\varphi _{\gamma }, & \left[ l^{z}=0\right] &  \\
\Psi _{R}^{\uparrow }(x,\mathbf{k}_{\perp },\downarrow ,\downarrow )=0, &  &
\end{array}
\right.
\end{equation}
in which:
\begin{equation}
\varphi _{\gamma }=\frac{e_{q}}{D}=\frac{e_{q}}{\lambda
^{2}-\frac{m^{2}+ \mathbf{k}_{\perp
}^{2}}{x}-\frac{m^{2}+\mathbf{k}_{\perp }^{2}}{1-x}},
\end{equation}%
where $\lambda $ is the photon mass and equals to 0. Each configuration
satisfies the spin sum rule:$J^{z}=S_{q}^{z}+S_{\overline{q}}^{z}+l^{z}=+1$.
Therefore, the quark-antiquark Fock-state for the photon $(J^{z}=+1)$ has
the four possible spin combinations:
\begin{eqnarray}
\left\vert \Psi _{\gamma }^{\uparrow }\left(
P^{+},\mathbf{P}_{\perp }\right) \right\rangle &=&\int
\frac{\mathrm{d}^{2}\mathbf{k}_{\perp }
\mathrm{d}x}{16{\pi }^{3}}  \nonumber \\
&&\times \left[ \Psi _{R}^{\uparrow }(x,\mathbf{k}_{\perp },\uparrow
,\downarrow )\left\vert xP^{+},\mathbf{k}_{\perp },\uparrow ,\downarrow
\right\rangle +\Psi _{R}^{\uparrow }(x,\mathbf{k}_{\perp },\downarrow
,\uparrow )\left\vert xP^{+},\mathbf{k}_{\perp },\downarrow ,\uparrow
\right\rangle \right.  \nonumber \\
&&\left. +\Psi _{R}^{^{\uparrow }}(x,\mathbf{k}_{\perp },\uparrow
,\uparrow )\left\vert xP^{+},\mathbf{k}_{\perp },\uparrow
,\uparrow \right\rangle +\Psi _{R}^{^{\uparrow
}}(x,\mathbf{k}_{\perp },\downarrow ,\downarrow )\left\vert
xP^{+},\mathbf{k}_{\perp },\downarrow ,\downarrow \right\rangle
\right] .  \nonumber \\
&&
\end{eqnarray}

The quark-antiquark Fock-state wave function of the pion is also
derived \cite{Xiao03} by using the relativistic field theory
treatment of the interaction vertex along with the idea in
\cite{BD80,Bro2001}. In the light-cone frame of pion,
\begin{equation}
\left\{
\begin{array}{lll}
P & = & (P^{+},\frac{M^{2}}{P^{+}},\mathbf{0}_{\perp }), \\
p_{1} & = & (xP^{+},\frac{\mathbf{p}_{1\perp
}^{2}+m^{2}}{xP^{+}},\mathbf{p}
_{1\perp }) \\
p_{2} & = & ((1-x)P^{+},\frac{\mathbf{p}_{2\perp
}^{2}+m^{2}}{(1-x)P^{+}}, \mathbf{p}_{2\perp }),
\end{array}
\right.
\end{equation}
we can obtain the four components of the spin wave function by calculating
the matrix elements of
\begin{equation}
\frac{\overline{v}(p_{2}^{+},p_{2}^{-},-\mathbf{k}_{\perp
})}{\sqrt{p_{2}^{+} }}\gamma
_{5}\frac{u(p_{1}^{+},p_{1}^{-},\mathbf{k}_{\perp })}{\sqrt{
p_{1}^{+}}},
\end{equation}%
from which we have
\begin{equation}
\left\{
\begin{array}{lll}
\frac{\overline{v}_{\downarrow }}{\sqrt{p_{2}^{+}}}\gamma
_{5}\frac{
u_{\uparrow }}{\sqrt{p_{1}^{+}}} & = & -\frac{2mP^{+}}{4mx(1-x)P^{+2}}, \\
\frac{\overline{v}_{\downarrow }}{\sqrt{p_{2}^{+}}}\gamma
_{5}\frac{
u_{\uparrow }}{\sqrt{p_{1}^{+}}} & = & +\frac{2mP^{+}}{4mx(1-x)P^{+2}}, \\
\frac{\overline{v}_{\uparrow }}{\sqrt{p_{2}^{+}}}\gamma _{5}\frac{
u_{\uparrow }}{\sqrt{p_{1}^{+}}} & = &
+\frac{2(k_{1}+ik_{2})P^{+}}{
4mx(1-x)P^{+2}}, \\
\frac{\overline{v}_{\downarrow }}{\sqrt{p_{2}^{+}}}\gamma
_{5}\frac{ u_{\downarrow }}{\sqrt{p_{1}^{+}}} & = &
+\frac{2(k_{1}-ik_{2})P^{+}}{ 4mx(1-x)P^{+2}},
\end{array}
\right.
\end{equation}
where $m$ is the mass of the quark. After the normalization, we
can obtain light-cone representation for the spin structure of the
pion, which is the minimal Fock-state of the pion light-cone wave
function:

\begin{equation}
\left\{
\begin{array}{lll}
\Psi _{\pi L}(x,\mathbf{k}_{\perp },\uparrow ,\downarrow
)=-\frac{m}{\sqrt{ 2(m^{2}+\mathbf{k}_{\perp }^{2})}}\varphi _{\pi
}, & \left[ l^{z}=0\right]
&  \\
\Psi _{\pi L}(x,\mathbf{k}_{\perp },\downarrow ,\uparrow
)=+\frac{m}{\sqrt{ 2(m^{2}+\mathbf{k}_{\perp }^{2})}}\varphi _{\pi
}, & \left[ l^{z}=0\right]
&  \\
\Psi _{\pi L}(x,\mathbf{k}_{\perp },\uparrow ,\uparrow
)=+\frac{k_{1}+ik_{2} }{\sqrt{2(m^{2}+\mathbf{k}_{\perp
}^{2})}}\varphi _{\pi }, & \left[ l^{z}=-1
\right]  &  \\
\Psi _{\pi L}(x,\mathbf{k}_{\perp },\downarrow ,\downarrow
)=+\frac{ k_{1}-ik_{2}}{\sqrt{2(m^{2}+\mathbf{k}_{\perp
}^{2})}}\varphi _{\pi }, & \left[ l^{z}=+1\right]  &
\end{array}
\right.   \label{LCpionWF}
\end{equation}
in which we may employ the Brodsky-Huang-Lepage (BHL) prescription
\cite{BHL81},
\begin{equation}
\varphi _{\pi }(x,\mathbf{k})=A\exp \left[ -\frac{1}{8{\beta
}^{2}}\frac{\mathbf{k}_{\perp }^{2}+m^{2}}{x(1-x)}\right] ,
\label{BHL}
\end{equation}
for the momentum space wave function, which is a non-relativistic
solution of the Bethe-Salpeter equation in an instantaneous
approximation in the rest frame for meson. Each configuration
satisfies the spin sum rule:
$J^{z}=S_{q}^{z}+S_{\overline{q}}^{z}+l^{z}=0$. Hence, the Fock
expansion of the two particle Fock-state for the pion has these
four possible spin combinations:
\begin{eqnarray}
\left\langle \Psi _{\pi }\left( P^{+},\mathbf{P}_{\perp
}=\mathbf{0}_{\perp }\right) \right\vert  & = & \int
\frac{\mathrm{d}^{2}\mathbf{k}_{\perp }
\mathrm{d}x}{16{\ \pi }^{3}} \nonumber\\
&  & \times \left[ \Psi _{\pi L}(x,\mathbf{k}_{\perp },\uparrow ,\downarrow
)\left\langle xP^{+},\mathbf{k}_{\perp },\uparrow ,\downarrow \right\vert
+\Psi _{\pi L}(x,\mathbf{k}_{\perp },\downarrow ,\uparrow )\left\langle
xP^{+},\mathbf{k}_{\perp },\downarrow ,\uparrow \right\vert \right.  \nonumber\\
&  & \left. +\Psi _{\pi L}(x,\mathbf{k}_{\perp },\uparrow ,\uparrow
)\left\langle xP^{+},\mathbf{k}_{\perp },\uparrow ,\uparrow \right\vert
+\Psi _{\pi L}(x,\mathbf{k}_{\perp },\downarrow ,\downarrow )\left\langle
xP^{+},\mathbf{k}_{\perp },\downarrow ,\downarrow \right\vert \right]. \nonumber\\
&&
\label{wavefunction}
\end{eqnarray}
There are two higher helicity $(\lambda _{1}+\lambda _{2}=\pm 1)$
components in the expression of the light-cone spin wave function
of the pion besides the ordinary helicity $(\lambda _{1}+\lambda
_{2}=0)$ components. Such higher helicity components come from the
Melosh-Wigner rotation in the light-cone quark model
\cite{Ma93,Xiao2002}, and the same effect plays an important role
to understand the proton \textquotedblleft spin puzzle" in the
nucleon case \cite{Ma91,Ma96}. One may also state that these
higher helicity components contain contribution from orbital
angular moment from a relativistic viewpoint \cite{MS98}.

In addition, we would like to add some more remarks on the
Gaussian-type wavefunction of the BHL prescription that we employ
above. As a matter of fact, the Gaussian wave function is a
non-relativistic solution of the Bethe-Salpeter equation in an
instantaneous approximation in the rest frame of the meson as the
space wave function.  The BHL wavefunction Eq.~(\ref{BHL}) is an
extension from a non-relativistic wavefunction into a relativistic
form by using the Brodsky-Huang-Lepage Ansatz~\cite{BHL81}, and we
can consider it as an approximate wavefunction that respects the
Lorentz invariance in the light-cone formalism. However, it works
phenomenologically well in a lot of calculations ($\textit{e.g.,}$
\cite{Hwang2001,Xiao2002,Xiao03,Don2001}). Moreover, Donnachie,
Gravelis, and Shaw \cite{Don2001} indicated that the other four
possible different space wave functions have similar analytical
properties with the BHL wavefunction when the parameter $\beta$ is
small (The $\beta$ is equal to $P_{F}$ in their paper, the small
$\beta$ is corresponding to the $\rho$ and $\phi$ mesons cases).
However, they also illustrated that the BHL wavefunction is better
than the other four wave functions in the high $\beta$ situation
(for the $J/\psi$ meson). Hence, it gives us the idea that the the
BHL wavefunction may be an appropriate choice that we could have
right now. (Noticing that the space wave functions for the vector
mesons are the same with those for the pseudoscalar mesons, the
argument that made by Donnachie $\textit{et al}$ is also valid for
$\pi ^{0}$ and other pseudoscalar mesons.)


For the physical state of $\pi ^{0}$, one should also take into account the
color and flavor degrees of freedom into account \cite{Lep80,Cao96}
\begin{equation}
\left\vert \Psi _{\pi ^{0}}\right\rangle =\sum_{a}\frac{\delta
_{b}^{a}}{ \sqrt{n_{c}}}\frac{1}{\sqrt{2}}\left[ \left\vert
u^{a}\bar{u} ^{b}\right\rangle -\left\vert
d^{a}\bar{d}^{b}\right\rangle \right] ,
\end{equation}%
where $a$ and $b$ are color indices, $n_{c}=3$ is the number of
colors, and now $\left\vert q^{a}\bar{q}^{b}\right\rangle $
contains the full spin structure shown above. So we can get the
photon-meson transition form factor of the pion:
\begin{eqnarray}
F_{\gamma \gamma ^{\ast }\rightarrow \pi }(Q^{2}) &=&\frac{\Gamma
^{+}}{ -ie^{2}(\mathbf{\epsilon }_{\perp }\times \mathbf{q}_{\perp
})p_{\pi
}^{\prime -}}  \nonumber \\
&=&2\sqrt{3}(e_{u}^{2}-e_{d}^{2})\int_{0}^{1}\mathrm{d}x\int
\frac{\mathrm{d} ^{2}\mathbf{k}_{\perp }}{16{\pi }^{3}}\varphi
_{\pi }(x,\mathbf{k}_{\perp
}^{\prime }) \nonumber\\
&&\left\{ \frac{m}{x\sqrt{m^{2}+\mathbf{k}_{\perp }^{\prime
2}}}\times \left[ \frac{1}{-\lambda
^{2}+\frac{m^{2}+\mathbf{k}_{\perp }^{2}}{x}+\frac{m^{2}+
\mathbf{k}_{\perp }^{2}}{1-x}}\right] +(1\leftrightarrow 2)\right\} \\
&=&4\sqrt{3}(e_{u}^{2}-e_{d}^{2})\int_{0}^{1}\mathrm{d}x\int
\frac{\mathrm{d} ^{2}\mathbf{k}_{\perp }}{16{\pi }^{3}}\left[
\varphi _{\pi }(x,\mathbf{k} _{\perp }^{\prime
})\frac{m}{x\sqrt{m^{2}+\mathbf{k}_{\perp }^{\prime 2}}}
\right.  \nonumber \\
&&\left. \times \frac{1}{\frac{m^{2}+\mathbf{k}_{\perp
}^{2}}{x}-\frac{m^{2}+ \mathbf{k}_{\perp }^{2}}{1-x}}\right] ,
\label{Fphotonpion}
\end{eqnarray}%
in which $\mathbf{k}_{\perp }^{\prime }=\mathbf{k}_{\perp
}+(1-x)\mathbf{q}_{\perp }$ after considering the Drell-Yan-West
assignment \cite{DYW}, and $\lambda $ $(=0)$ is the mass of
photon.

\section{The $\protect\eta \mbox{-}\protect\eta ^{\prime }$ mixing schemes}

In fact, there are two popular mixing schemes for $\eta $ and
$\eta ^{\prime }$. Feldmann \textit{et al }\cite{Feld99} suggested
the mixing scheme based on the quark flavor basis $q\overline{q}=(u\overline{u}+d\overline{d})/%
\sqrt{2}$ and $s\overline{s}$,
\begin{equation}
\left(
\begin{array}{c}
\eta \\
\eta ^{\prime }%
\end{array}
\right) =\left(
\begin{array}{cc}
\cos \phi & -\sin \phi \\
\sin \phi & \cos \phi%
\end{array}
\right) \left(
\begin{array}{c}
\eta _{q} \\
\eta _{s}%
\end{array}
\right),  \label{Mixing}
\end{equation}
and
\begin{equation}
\left(
\begin{array}{cc}
f_{\eta }^{q} & f_{\eta }^{s} \\
f_{\eta \prime }^{q} & f_{\eta \prime }^{s}%
\end{array}
\right) =\left(
\begin{array}{cc}
\cos \phi & -\sin \phi \\
\sin \phi & \cos \phi%
\end{array}
\right) \left(
\begin{array}{cc}
f_{q} & 0 \\
0 & f_{s}%
\end{array}
\right),
\end{equation}
where $\phi $ is the mixing angle. The
$q\overline{q}\mbox{-}s\overline{s}$ mixing only introduces one
mixing angle in the mixing of the decay constants.

On the other hand, people also use the mixing scheme based on the
basis of $\eta _{8}$\ and $\eta _{0}$ mixing for $\eta $ and $\eta
^{\prime }$. In the $SU(3)$ quark model, the physical $\eta $ and
$\eta ^{\prime }$ states dominantly consist of a flavor $SU(3)$
octet $\eta _{8}=\frac{1}{\sqrt{6}}(u\overline{u}+d
\overline{d}-2s\overline{s})$ and a singlet $\eta _{0}=$
$\frac{1}{ \sqrt{3}}(u\overline{u}+d\overline{d}+s\overline{s})$,
respectively. The usual mixing scheme reads:
\begin{equation}
\left\{
\begin{array}{lll}
\left\vert \eta \right\rangle &=&\cos \theta \left\vert \eta
_{8}\right\rangle -\sin \theta \left\vert \eta _{0}\right\rangle,

\\
\left\vert \eta ^{\prime }\right\rangle &=&\sin \theta \left\vert
\eta _{8}\right\rangle +\cos \theta \left\vert \eta
_{0}\right\rangle,
\end{array}
\right.
\end{equation}
in which $\theta $ is the mixing angle. For the calculation of the
decay constants of the $\eta _{8}$\ and $\eta _{0}$ mixing,
Feldmann-Kroll indicate that two mixing angle scheme could be
better from their former investigations\cite{Feld98}:

\begin{equation}
\left(
\begin{array}{cc}
f_{\eta }^{8} & f_{\eta }^{0} \\
f_{\eta \prime }^{8} & f_{\eta \prime }^{0}%
\end{array}%
\right) =\left(
\begin{array}{cc}
\cos \theta _{8} & -\sin \theta _{0} \\
\sin \theta _{8} & \cos \theta _{0}%
\end{array}%
\right) \left(
\begin{array}{cc}
f_{8} & 0 \\
0 & f_{0}%
\end{array}%
\right).
\end{equation}

In addition, one could find that these two schemes could be
related by the following equation through the mixing angles
finally:
\begin{equation}
\theta =\phi -\arctan \frac{1}{\sqrt{3}}.
\end{equation}

From the point view of the chiral symmetry and the $SU(3)$
symmetry as well as their breaking mechanisms, we find that the
$\eta _{8}$\ and $\eta _{0}$ mixing scheme may be more reasonable
and physical.

First of all, let us have a brief review on the chiral symmetry
and its breaking which have underlying relationship with the $\pi
^{0}$, $\eta $ and $\eta ^{\prime }$ mesons\cite{Gan}. In the
chiral symmetry limit, it is well-known that the Lagrangian has
the $SU(3)_{L}\times SU(3)_{R}\times U(1)_{B}\times U(1)_{A}$
symmetry, but the absence of this symmetry in the ground state
(the QCD vacuum) leads to the chiral symmetry spontaneously
breaking into $SU(3)\times U(1)_{B}$ symmetry. Because there are 8
spontaneously broken continuous symmetries (there are 9 when
taking into account the chiral anomaly which is associated with
the the $U(1)_{A}$ symmetry breaking), there are 8 massless
Goldstone Bosons (which finally are identified as meson octet) and
1 massive particle (which is known as $\eta _{0}$) according to
the Goldstone's theorem and chiral anomaly, respectively. The
massless octet includes the meson $\pi _{8}^{0}$ and $\eta _{8}$.
Together with $\eta _{0}$, they mix into massive mesons $\pi
^{0}$, $\eta $ and $\eta ^{\prime }$ during the explicit $SU(3)$
symmetry breaking after introducing the quark mass term into the
Lagrangian.

From the above discussion, we may reach a physical intuitive idea
that it is natural and straightforward to use the $\eta _{0}$ and
$\eta _{8}$ mixing scheme as a direct result of the $SU(3)$
symmetry breaking if we assume that the $\pi _{8}^{0}$ does not
mix with $\eta _{0}$ and $\eta _{8}$ at all. From this
point of view, the introduction of $\eta _{0}$ and $\eta _{8}$ is more reasonable than $%
\eta _{q}$ and $\eta _{s}$.

Moreover, since it is well-known that pion, kaon, and $\eta _{8}$
belong to the same group of octet mesons in the $SU(3)$ symmetry
limit, their parameters should be the same except the quark
masses. In this sense, one may relate the decay constants of $\eta
$ and $\eta ^{\prime }$, to pion and kaon in the $\eta _{8}-\eta
_{0}$ mixing scheme. The CLEO Collaboration\cite{CLEO98} reported
their pole fit results as $\Lambda _{\pi }=776\pm 10\pm12\pm
16~MeV$, $\Lambda _{\eta }=774\pm 11\pm 16\pm 22~MeV$, and
$\Lambda _{\eta \prime }=859\pm 9\pm 18\pm 20~MeV$. These results
imply that the nonperturbative properties of $\pi $ and $\eta $
are very similar. In addition, the absolute value of $\theta $ is
small and $\cos \theta \left\vert \eta _{8}\right\rangle $ is the
leading order in the $\eta_{8}-\eta _{0}$ mixing scheme of the
$\eta $. They are consistent with the basic physical intuition
that both $\pi $ and $\eta _{8}$ are in the $SU_{f}(3)$ octet and
are pseudo-massless Goldstone particles. Therefore, that is why
the authors of \cite{Cao98} take the parameters of $\eta _{8}$ as
equal to pion, such as $b_{8}=b_{\pi }$ in their paper. From a
strict sense, if pion, kaon, and $\eta _{8}$ are in the same group
of octet mesons, the mass of $m_{q}$, $m_{s} $, and $\beta
_{8}=\beta _{\pi }$ in the BHL wave function should be the same in
the calculations of the $\pi $, $\eta $ and $\eta ^{\prime }$
transition form factors.

Therefore, we employ the intuitive $\eta _{8}\mbox{-}\eta  _{0}$
mixing scheme in the calculations of the $\pi $, $\eta $ and $\eta
^{\prime }$ transition form factors by using the uniform
parameters, which shows that the $SU(3)$ symmetry limit works well
in this work.

In practice, we utilize the $SU_{f}(3)$ broken form of wave
functions for flavor octet $\eta _{8}$ and singlet $\eta _{0}$:
\begin{eqnarray}
\left\vert \eta _{8}\right\rangle
&=&\frac{1}{\sqrt{6}}(u\overline{u}+d \overline{d})\phi
_{8}^{q}(x,\mathbf{k}_{\perp })-\frac{2}{\sqrt{6}}s
\overline{s}\phi _{8}^{s}(x,\mathbf{k}_{\perp }),  \label{eta8} \\
\left\vert \eta _{0}\right\rangle
&=&\frac{1}{\sqrt{3}}(u\overline{u}+d \overline{d})\phi
_{0}^{q}(x,\mathbf{k}_{\perp })+\frac{1}{\sqrt{3}}s
\overline{s}\phi _{0}^{s}(x,\mathbf{k}_{\perp }),  \label{eta0}
\end{eqnarray}%
in which we use Gaussian wave function of the BHL prescription:
\begin{eqnarray}
\phi _{8}^{q}(x,\mathbf{k}_{\perp }) &=&A_{8}\exp \left[
-\frac{m_{q}^{2}+
\mathbf{k}_{\perp }^{2}}{8\beta _{8}^{2}x(1-x)}\right] , \\
\phi _{8}^{s}(x,\mathbf{k}_{\perp }) &=&A_{8}\exp \left[
-\frac{m_{s}^{2}+
\mathbf{k}_{\perp }^{2}}{8\beta _{8}^{2}x(1-x)}\right] , \\
\phi _{0}^{q}(x,\mathbf{k}_{\perp }) &=&A_{0}\exp \left[
-\frac{m_{q}^{2}+
\mathbf{k}_{\perp }^{2}}{8\beta _{0}^{2}x(1-x)}\right] , \\
\phi _{0}^{s}(x,\mathbf{k}_{\perp }) &=&A_{0}\exp \left[
-\frac{m_{s}^{2}+ \mathbf{k}_{\perp }^{2}}{8\beta
_{0}^{2}x(1-x)}\right] ,
\end{eqnarray}%
and $q\overline{q}$ and $s\overline{s}$ are the spin parts of the
wave functions which are similar to the pion with all possible
spin states.

Moreover, it is convenient to use the method for the $\eta
_{8}\mbox{-}\eta _{0}$ mixing scheme which was developed by
Cao-Signal \cite{Cao99} in obtaining the mixing angle $\theta $
and the decay constants. In this treatment, we can get $\theta $,
$f_{8}$ and $f_{0}$ directly without involving $\theta _{8}$ and
$\theta _{0}$. In the $\eta _{8}\mbox{-}\eta _{0}$ mixing scheme,
we have:
\begin{eqnarray}
F_{\gamma \gamma ^{\ast }\rightarrow \eta }(Q^{2}) &=&F_{\gamma \gamma
^{\ast }\rightarrow \eta _{8}}(Q^{2})\cos \theta -F_{\gamma \gamma ^{\ast
}\rightarrow \eta _{_{0}}}(Q^{2})\sin \theta , \\
F_{\gamma \gamma ^{\ast }\rightarrow \eta \prime }(Q^{2}) &=&F_{\gamma
\gamma ^{\ast }\rightarrow \eta _{8}}(Q^{2})\sin \theta +F_{\gamma \gamma
^{\ast }\rightarrow \eta _{0}}(Q^{2})\cos \theta .
\end{eqnarray}
While for the $\pi ^{0}$ case, we have:
\begin{eqnarray}
\Gamma (\pi ^{0} &\rightarrow &\gamma \gamma )=\frac{\pi \alpha ^{2}m_{\pi
^{0}}^{3}}{4}|F_{\gamma \gamma ^{\ast }\rightarrow \pi }(0)|^{2}, \label{pi_gamma}\\
\Gamma (\pi ^{0} &\rightarrow &\gamma \gamma )=\frac{\alpha
^{2}m_{\pi ^{0}}^{3}}{64\pi ^{3}}\frac{1}{f_{\pi }^{2}}.
\end{eqnarray}
Generalizing these equations to $\eta _{8}$ and $\eta _{0}$, we
have
\begin{eqnarray}
\Gamma (\eta &\rightarrow &\gamma \gamma )=\frac{\pi \alpha ^{2}m_{\eta
}^{3} }{4}|F_{\gamma \gamma ^{\ast }\rightarrow \eta }(0)|^{2}=\frac{\alpha
^{2}m_{\eta }^{3}}{64\pi ^{3}}\left( \frac{\cos \theta }{\sqrt{3}f_{8}}-
\frac{2\sqrt{2}\sin \theta }{\sqrt{3}f_{0}}\right) ^{2}, \\
\Gamma (\eta ^{\prime } &\rightarrow &\gamma \gamma )=\frac{\pi \alpha
^{2}m_{\eta \prime }^{3}}{4}|F_{\gamma \gamma ^{\ast }\rightarrow \eta
\prime }(0)|^{2}=\frac{\alpha ^{2}m_{\eta ^{\prime }}^{3}}{64\pi ^{3}}\left(
\frac{\sin \theta }{\sqrt{3}f_{8}}+\frac{2\sqrt{2}\cos \theta }{\sqrt{3}%
f_{0} }\right) ^{2} \label{eta_gamma}.
\end{eqnarray}
Thus we could get:
\begin{eqnarray}
\rho _{1} &=&\frac{F_{\gamma \gamma ^{\ast }\rightarrow \eta }(0)}{F_{\gamma
\gamma ^{\ast }\rightarrow \eta \prime }(0)}=\frac{\tan \theta _{08}-\tan
\theta }{1+\tan \theta _{08}\times \tan \theta }, \\
&=&\tan \left( \theta _{08}-\theta \right) ,
\end{eqnarray}%
in which we let $\tan \theta _{08}=\frac{f_{0}}{\sqrt{8}f_{8}}$.
Along with the same idea by taking the $Q^{2}\rightarrow \infty $
limit, we could have:
\begin{eqnarray}
\rho _{2} &=&\frac{F_{\gamma \gamma ^{\ast }\rightarrow \eta
}(Q^{2}\rightarrow \infty )}{F_{\gamma \gamma ^{\ast }\rightarrow
\eta \prime }(Q^{2}\rightarrow \infty )}=\frac{F_{\gamma \gamma
^{\ast }\rightarrow \eta _{8}}(Q^{2}\rightarrow \infty )\cos
\theta -F_{\gamma \gamma ^{\ast }\rightarrow \eta
_{_{0}}}(Q^{2}\rightarrow \infty )\sin \theta }{F_{\gamma \gamma
^{\ast }\rightarrow \eta _{8}}(Q^{2}\rightarrow \infty )\sin
\theta +F_{\gamma \gamma ^{\ast }\rightarrow \eta
_{_{0}}}(Q^{2}\rightarrow \infty )\cos \theta } \nonumber \\
&=&\frac{1-8\tan \theta _{08}\times \tan \theta }{\tan \theta
+8\tan \theta _{08}},
\end{eqnarray}
in which we have $\lim_{Q^{2}\rightarrow \infty
}Q^{2}F_{8}(Q^{2})=\frac{ 2f_{8}}{\sqrt{3}}$ and
$\lim_{Q^{2}\rightarrow \infty }Q^{2}F_{0}(Q^{2})=
\frac{4\sqrt{2}f_{0}}{\sqrt{3}}$. CLEO \cite{CLEO98} proposed that
the $\gamma \gamma ^{\ast }\rightarrow M$ transition form factors
could be approximated by:
\begin{equation}
F_{\gamma \gamma ^{\ast }\rightarrow M}(Q^{2})=F_{\gamma \gamma
^{\ast }\rightarrow M}(0)\times \frac{1}{1+Q^{2}/\Lambda
_{M}^{2}},
\end{equation}
thus we obtain:
\begin{equation}
\rho _{2}=\rho _{1}\frac{\Lambda _{\eta }^{2}}{\Lambda _{\eta
\prime }^{2}}.
\end{equation}
Finally one obtains:
\begin{eqnarray}
\tan \theta &=&\frac{-9(\rho _{1}+\rho _{2})+\sqrt{81(\rho _{1}-\rho
_{2})^{2}+32(\rho _{1}\rho _{2}+1)^{2}}}{2(8-\rho _{1}\rho _{2})},
\label{mixing} \\
\frac{f_{0}}{f_{8}} &=&\sqrt{8}\tan (\theta +\arctan \rho _{1}),
\end{eqnarray}%
and gets
\begin{eqnarray}
f_{8} &=&\frac{1}{4\sqrt{3}\pi ^{2}\left[F_{\gamma \gamma ^{\ast
}\rightarrow \eta }(0)\cos \theta +F_{\gamma \gamma ^{\ast
}\rightarrow \eta \prime
}(0)\sin \theta \right]},  \\
f_{0} &=&\frac{\sqrt{8}}{4\sqrt{3}\pi ^{2}\left[F_{\gamma \gamma
^{\ast }\rightarrow \eta \prime }(0)\sin \theta -F_{\gamma \gamma
^{\ast }\rightarrow \eta }(0)\cos \theta \right]},
\label{decayconstant}
\end{eqnarray}
by using the above results.

\section{$\protect\gamma ^{\ast }\protect\gamma \rightarrow \protect\eta $
and $\protect\gamma ^{\ast }\protect\gamma \rightarrow \protect\eta $
transition form factors}

There have been many different approaches to discuss the
photon-meson transition form factors of light pseudoscalar mesons
$\pi^0$, $\eta$, and $\eta'$, such as the light-cone perturbation
theory by Cao-Huang-Ma \cite{Cao96,Cao98}, the light-front quark
model by Hwang and Choi-Ji \cite{Hwang2001}, QCD sum rule
calculation by Radyushkin-Ruskov \cite{Rad97}, and also other
approaches \textit{et al.} \cite{LitTran}. We now perform a
systematic calculation of these transition form factors in the
light-cone framework just presented in section 2. The advantage of
this new framework is that the predictions should be applicable at
both low and high energy scales.

Similar to the pion transition form factor and from
Eq.~(\ref{eta8}) and Eq.~(\ref{eta0}), we can get $\eta _{8}$ and
$\eta _{0}$ transition form factors:

\begin{eqnarray}
F_{\gamma \gamma ^{\ast }\rightarrow \eta _{8}}(Q^{2})
&=&4(e_{u}^{2}+e_{d}^{2})\int \frac{{\mathrm{d}}x{\mathrm{d}}^{2}\mathbf{k}
_{\perp }}{16{\pi }^{3}}\frac{m_{q}}{x\sqrt{m_{q}^{2}+\mathbf{k}_{\perp
}^{\prime 2}}}\phi _{8}^{q}(x,\mathbf{k}_{\perp }^{\prime })\frac{x(1-x)}{
m_{q}^{2}+\mathbf{k}_{\perp }^{2}}  \nonumber \\
&&-8e_{s}^{2}\int \frac{{\mathrm{d}}x{\mathrm{d}}^{2}\mathbf{k}_{\perp }}{16{%
\pi }^{3}}\frac{m_{s}}{x\sqrt{m_{s}^{2}+\mathbf{k}_{\perp }^{\prime 2}}}\phi
_{8}^{s}(x,\mathbf{k}_{\perp }^{\prime })\frac{x(1-x)}{m_{s}^{2}+\mathbf{k}
_{\perp }^{2}}, \\
F_{\gamma \gamma ^{\ast }\rightarrow \eta _{0}}(Q^{2}) &=&4\sqrt{2}
(e_{u}^{2}+e_{d}^{2})\int \frac{{\mathrm{d}}x{\mathrm{d}}^{2}\mathbf{k}
_{\perp }}{16{\pi }^{3}}\frac{m_{q}}{x\sqrt{m_{q}^{2}+\mathbf{k}_{\perp
}^{\prime 2}}}\phi _{0}^{q}(x,\mathbf{k}_{\perp }^{\prime })\frac{x(1-x)}{
m_{q}^{2}+\mathbf{k}_{\perp }^{2}} \nonumber\\
&&+4\sqrt{2}e_{s}^{2}\int
\frac{{\mathrm{d}}x{\mathrm{d}}^{2}\mathbf{k} _{\perp }}{16{\pi
}^{3}}\frac{m_{s}}{x\sqrt{m_{s}^{2}+\mathbf{k}_{\perp }^{\prime
2}}}\phi _{0}^{s}(x,\mathbf{k}_{\perp }^{\prime })\frac{x(1-x)}{
m_{s}^{2}+\mathbf{k}_{\perp }^{2}},
\end{eqnarray}
in which $\mathbf{k}_{\perp }^{\prime }=\mathbf{k}_{\perp
}+(1-x)\mathbf{q} _{\perp }$ after considering the Drell-Yan-West
assignment, and then we get $F_{\gamma \gamma ^{\ast }\rightarrow
\eta }(Q^{2})$ and $F_{\gamma \gamma ^{\ast }\rightarrow \eta
\prime }(Q^{2})$ in the $\eta _{8}$-$\eta _{0}$ mixing scheme
\begin{eqnarray}
F_{\gamma \gamma ^{\ast }\rightarrow \eta }(Q^{2}) &=&F_{\gamma \gamma
^{\ast }\rightarrow \eta _{8}}(Q^{2})\cos \theta -F_{\gamma \gamma ^{\ast
}\rightarrow \eta _{0}}(Q^{2})\sin \theta , \\
F_{\gamma \gamma ^{\ast }\rightarrow \eta \prime }(Q^{2}) &=&F_{\gamma
\gamma ^{\ast }\rightarrow \eta _{8}}(Q^{2})\sin \theta +F_{\gamma \gamma
^{\ast }\rightarrow \eta _{0}}(Q^{2})\cos \theta .
\end{eqnarray}

\subsection{Numerical calculations}

First of all, we would like to determine the mixing angle $\theta
$ and decay constants of $\ f_{8}$ and $f_{0}$ by employing
Eq.~(\ref{mixing}) to Eq.~(\ref{decayconstant}) with two different
sets of experimental data which may cast some light on the
clarification of the obvious current disagreement between the
former Primakoff experiments and collider results in the
measurements of $\Gamma (\eta \rightarrow \gamma \gamma )$, and
then give more reasonable predictions on the mixing angle $\theta
$. From the Particle Data Group book \cite{PDB}, we get:%
\begin{eqnarray}
\Gamma (\pi ^{0} &\rightarrow &\gamma \gamma )=7.74\pm 0.54~\mbox{eV}, \\
\Gamma (\eta &\rightarrow &\gamma \gamma )=0.46\pm 0.04~\mbox{keV}, \\
\Gamma (\eta ^{\prime } &\rightarrow &\gamma \gamma )=4.29\pm
0.15~\mbox{keV},
\end{eqnarray}%
and%
\begin{eqnarray}
m_{\pi ^{0}} &=&134.9766\pm 0.0006~\mbox{MeV}, \\
m_{\eta } &=&547.30\pm 0.12~\mbox{MeV}, \\
m_{\eta \prime } &=&957.78\pm 0.14~\mbox{MeV}.
\end{eqnarray}%
We can get $\theta =-14.7^{\circ }\pm 2.0^{\circ },$
$f_{0}=1.13\pm 0.08f_{\pi }$ and $f_{8}=0.97\pm 0.07f_{\pi }$.
However, $\Gamma (\eta \rightarrow \gamma \gamma )=0.511\pm
0.026~\mbox{keV}$ if we do not include the
Primakoff production measurement of $\eta \rightarrow \gamma \gamma $ $%
(\Gamma (\eta \rightarrow \gamma \gamma )=0.324\pm
0.046~\mbox{keV})$ which obviously disagrees with other collider
measurement. Therefore, we obtain $\theta =-16.1^{\circ }\pm
1.5^{\circ },$ $f_{0}=1.11\pm 0.08f_{\pi }$
and $f_{8}=0.95\pm 0.07f_{\pi }$. Moreover, we find that the mixing angle $%
\phi=\theta+\arctan \frac{1}{\sqrt{3}}=38.6^{\circ}$ is compatible with \cite%
{Feld99} which gives the phenomenological value of the mixing angle $%
\phi=39.3^{\circ }\pm 1.0^{\circ}$ from eight decay and scattering
processes. The mixing independent ratio $R$ can be defined as
follow:
\begin{eqnarray}
R &\equiv &\frac{M_{\pi }^{3}}{\Gamma (\pi \rightarrow \gamma \gamma )}\left[
\frac{\Gamma (\eta \rightarrow \gamma \gamma )}{M_{\eta }^{3}}+\frac{\Gamma
(\eta ^{\prime }\rightarrow \gamma \gamma )}{M_{\eta \prime }^{3}}\right]
\label{r1} \\
&=&\frac{1}{3}\left( \frac{f_{\pi }^{2}}{f_{8}^{2}}+8\frac{f_{\pi }^{2}}{%
f_{0}^{2}}\right) .
\end{eqnarray}
The current experimental value of $R$ which was given in the
proposal of the PrimEx Collaboration \cite{Gan} at JLab  is
$R_{\exp }=2.5\pm 0.5(\mbox{stat})\pm 0.5(\mbox{syst})$. We can
get $R=2.45$ and $R=2.54$ respectively by using the above two sets
of the parameters. With the latter set of the fitted value of the
mixing angle $\theta $ and decay constants of $\ f_{8}$ and
$f_{0}$ as the input parameters, we can fix the left seven
parameters by the following nine constraints.

In the formulas of the transition form factor $F_{\gamma \gamma
^{\ast }\rightarrow P}(Q^{2})$ $(P=\pi ,\eta _{8},\eta _{0})$, the
parameters are the
normalization constants $A_{\pi }$, $A_{8}$ and $A_{0}$, the harmonic scale $%
\beta _{\pi }=\beta _{8}$ and $\beta _{0}$, and the quark masses $%
m_{q}=m_{u}=m_{d}$ and $m_{s}$. In order to take a numerical
calculation of the transition form factor $F_{\gamma \gamma ^{\ast
}\rightarrow M}(Q^{2})$ and compare it with the available
experimental data, we need to employ nine constraints to fix those
seven parameters above. Thus, we can determine all these seven
parameters in the transition form factor uniquely.

\noindent
1.~The decay widths of $\pi $, $\eta $ and $\eta ^{\prime }$ \cite{CLEO98,PDB}%
:
\begin{eqnarray}
F_{\pi \gamma }(0) &=&\sqrt{\frac{4}{\alpha ^{2}\pi M_{\pi
}^{3}}\Gamma (\pi \rightarrow \gamma \gamma )}=0.274\pm
0.010~\mbox{GeV}^{-1},\ \ 0.274~\mbox{GeV}^{-1},
\label{constraints1}\\
F_{\eta \gamma }(0) &=&\sqrt{\frac{4}{\alpha
^{2}\pi M_{\eta }^{3}}\Gamma (\eta \rightarrow \gamma \gamma
)}=0.273\pm 0.009~\mbox{GeV}^{-1},\ \ 0.277~\mbox{GeV}^{-1},
\label{constraints2}\\
F_{\eta \prime \gamma }(0) &=&\sqrt{\frac{4}{\alpha ^{2}\pi M_{\eta
\prime }^{3}}\Gamma (\eta ^{\prime }\rightarrow \gamma \gamma
)}=0.342\pm 0.006~\mbox{GeV}^{-1},\ \
0.343~\mbox{GeV}^{-1}.\label{constraints3}
\end{eqnarray}

\noindent 2.~The $Q^{2}\rightarrow \infty $ limiting behavior of $%
Q^{2}F_{\gamma \gamma ^{\ast }\rightarrow P}(0)F_{\gamma \gamma
^{\ast }\rightarrow P}(Q^{2})$ \cite{Lep80,Cao99,Bro81}:
\begin{eqnarray}
\lim_{Q^{2}\rightarrow \infty }\pi ^{2}Q^{2}F_{\gamma \gamma
^{\ast }\rightarrow \pi }(0)F_{\gamma \gamma ^{\ast }\rightarrow
\pi }(Q^{2}) &=& \frac{1}{2},\ \ 0.49, \label{constraints4}
\\
\lim_{Q^{2}\rightarrow \infty }3\pi ^{2}Q^{2}F_{\gamma \gamma ^{\ast
}\rightarrow \eta _{8}}(0)F_{\gamma \gamma ^{\ast }\rightarrow \eta
_{8}}(Q^{2}) &=&\frac{1}{2},\ \ 0.48, \label{constraints5} \\
\lim_{Q^{2}\rightarrow \infty }3\pi ^{2}Q^{2}F_{\gamma \gamma
^{\ast }\rightarrow \eta _{0}}(0)F_{\gamma \gamma ^{\ast
}\rightarrow \eta _{0}}(Q^{2}) &=&4, \label{constraints6} \ \
3.99.
\end{eqnarray}

\noindent 3.~The $Q^{2}\rightarrow \infty $ limiting behavior of $%
Q^{2}F_{\gamma \gamma ^{\ast }\rightarrow P}(Q^{2})$
\cite{Lep80,Cao99,Bro81}:
\begin{eqnarray}
\lim_{Q^{2}\rightarrow \infty }Q^{2}F_{\gamma \gamma ^{\ast }\rightarrow \pi
}(Q^{2}) &=&2f_{\pi }=184.8\pm 0.2~\mbox{MeV}, \ \ 184.8~\mbox{MeV},\label{constraints7} \\
\lim_{Q^{2}\rightarrow \infty }Q^{2}F_{\gamma \gamma ^{\ast
}\rightarrow
\eta _{8}}(Q^{2}) &=&\frac{2}{\sqrt{3}}f_{8}=101\pm 7~\mbox{MeV},\ \  95~\mbox{MeV},\label{constraints8} \\
\lim_{Q^{2}\rightarrow \infty }Q^{2}F_{\gamma \gamma ^{\ast
}\rightarrow \eta _{0}}(Q^{2})
&=&\frac{4\sqrt{2}}{\sqrt{3}}f_{0}=334\pm 15~\mbox{MeV},\ \
332~\mbox{MeV},\label{constraints9}
\end{eqnarray}%
in which the weak decay constant $f_{\pi }=92.4~\mbox{MeV}$ is
defined \cite{decay} from $\pi \rightarrow \mu \nu $ decay.

These constraints are not completely independent, but are
necessary since some of them are free from uncertainties, for
example, Eqs.~(\ref{constraints5}-\ref{constraints6}) are free
from the decay constants $f_{0}$ and $f_{8}$. Combined with
consideration of other properties of the pion \cite{Xiao03}, we
can obtain $m_{q}=200~\mbox{MeV}$, $m_{s}=550~\mbox{MeV}$, $\beta
_{\pi
}=\beta _{8}=410~\mbox{MeV}$, $\beta _{0}=475~\mbox{MeV}$, $A_{\pi }=0.0475~\mbox{MeV}^{-1}$, $%
A_{8}=0.0331~\mbox{MeV}^{-1}$, and $A_{0}=0.0440~\mbox{MeV}^{-1}$.
Among these 7 parameters, 3 of them ($m_{q}$, $A_{\pi}$ and $\beta
_{\pi}$) are the same in our previous work \cite{Xiao03} and have
already been fixed, only the other 4 are new parameters. These
three parameters satisfy Eqs.~(\ref{constraints1}),
(\ref{constraints4}) and (\ref{constraints7}) very well. Then we
fix the 4 new parameters by using the four equations
Eqs.~(\ref{constraints2}), (\ref{constraints3}),
(\ref{constraints5}) and (\ref{constraints6}). Since the parameter
fixing scheme is somehow unique, numerical results of these
parameters do not have much room to vary, and not surprisingly we
find these fixed 7 parameters give very good prediction for
Eqs.~(\ref{constraints8})-(\ref{constraints9}). Reversely, we can
compute the values of the above nine constraints by using the
above seven fixed parameters, and we also provide the fitted
values at the end of each equation. Therefore, after this simple
parameter fixing scheme, we could start to calculate the
transition form factor for these mesons.

\begin{figure}
\par
\begin{center}
\includegraphics{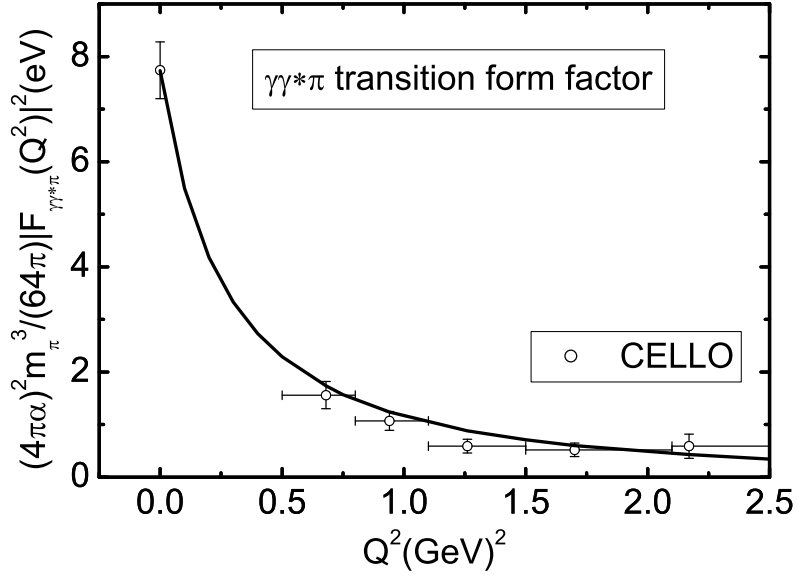}
\end{center}
\caption[*]{\baselineskip13pt Theoretical prediction of $(4
\protect\pi \protect\alpha )^{2}\frac{m_{\protect\pi
}^{3}}{64\protect\pi }
|F_{\protect\gamma \protect\gamma ^{\ast }\rightarrow \protect\pi %
}(Q^{2})|^{2}$ calculated with the pion wave function in the BHL
prescription compared with the experimental data. The data for the
transition form factor are taken from Ref.~\protect\cite{Beh91}.}
\label{pi1}
\end{figure}

\begin{figure}
\par
\begin{center}
\includegraphics{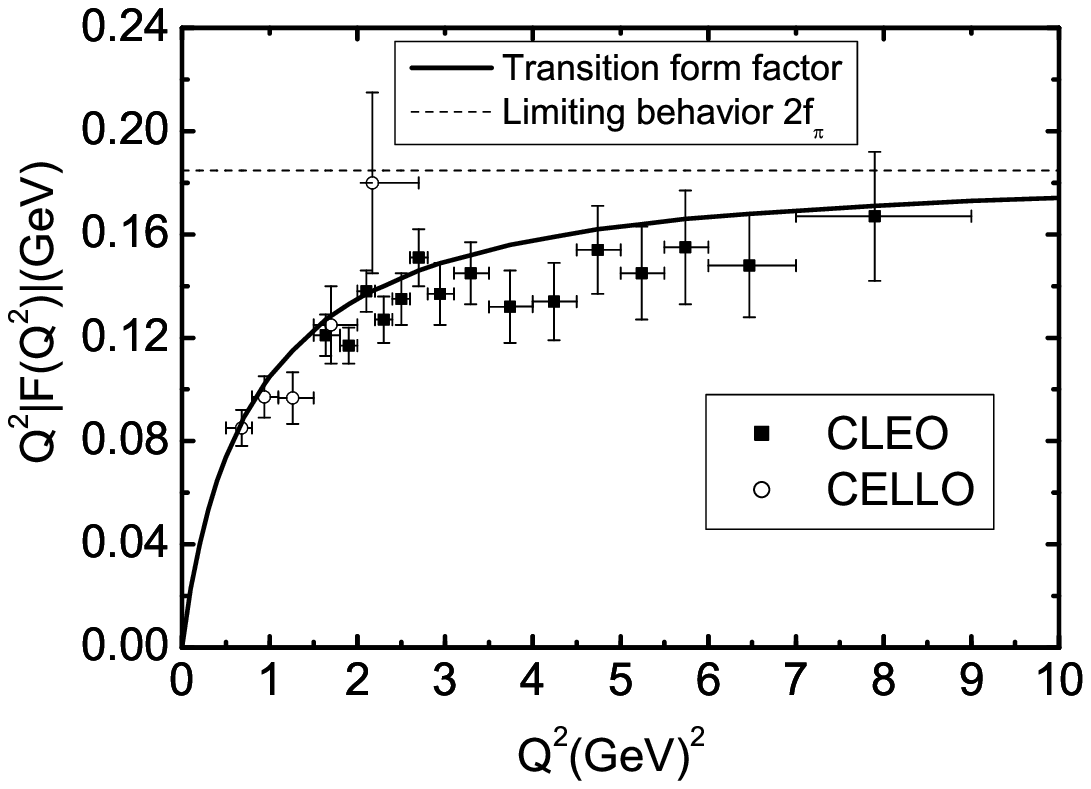}
\end{center}
\caption[*]{\baselineskip13pt Theoretical prediction  of $
Q^{2}|F_{\protect\gamma \protect\gamma ^{\ast }\rightarrow
\protect\pi}(Q^{2})|$ calculated with the pion wave function in
the BHL prescription compared with the experimental data. The data
for the transition form factor are taken from
Ref.~\protect\cite{CLEO98} and Ref.~\protect\cite{Beh91}.}
\label{pi2}
\end{figure}

\begin{figure}
\par
\begin{center}
\includegraphics{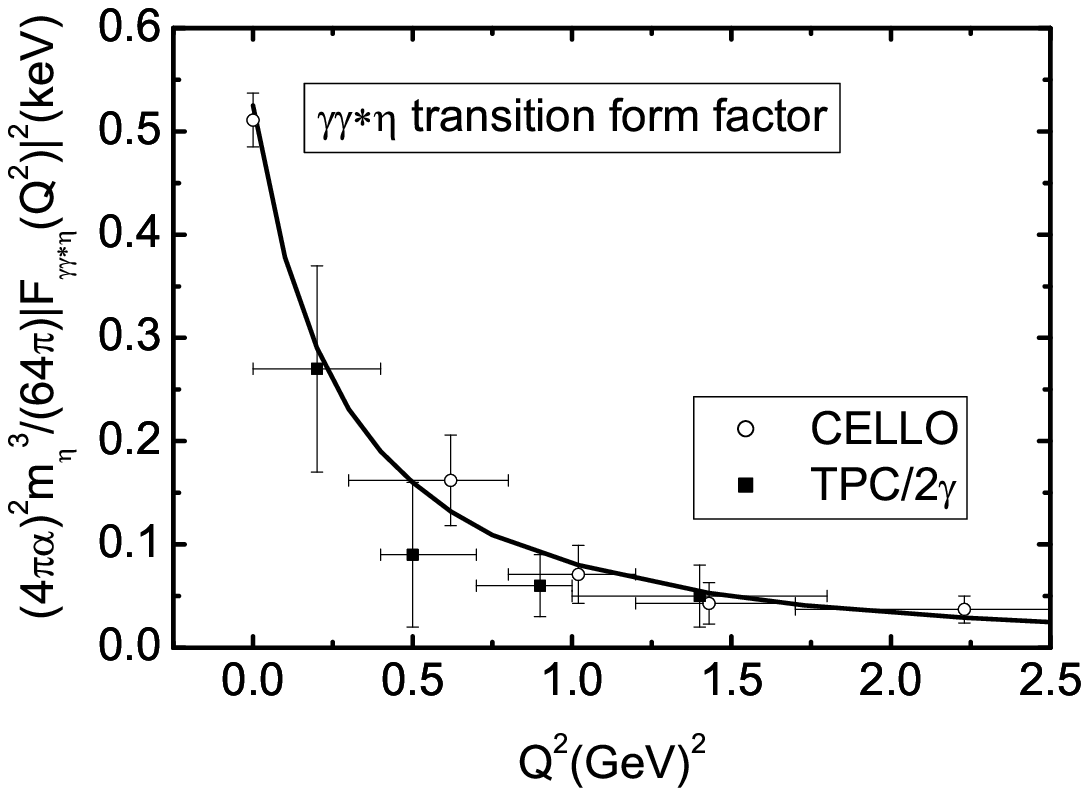}
\end{center}
\caption[*]{\baselineskip13pt Theoretical prediction of $(4
\protect\pi \protect\alpha )^{2}\frac{m_{\protect\eta
}^{3}}{64\protect\pi} |F_{\protect\gamma \protect\gamma ^{\ast
}\rightarrow \protect\eta }(Q^{2})|^{2}$ compared with the
experimental data in the low energy scale. The data for the
transition form factor are taken from Ref.~\protect\cite{Beh91}
and Ref.~\protect\cite{TPC90}.} \label{eta1}
\end{figure}

\begin{figure}
\par
\begin{center}
\includegraphics{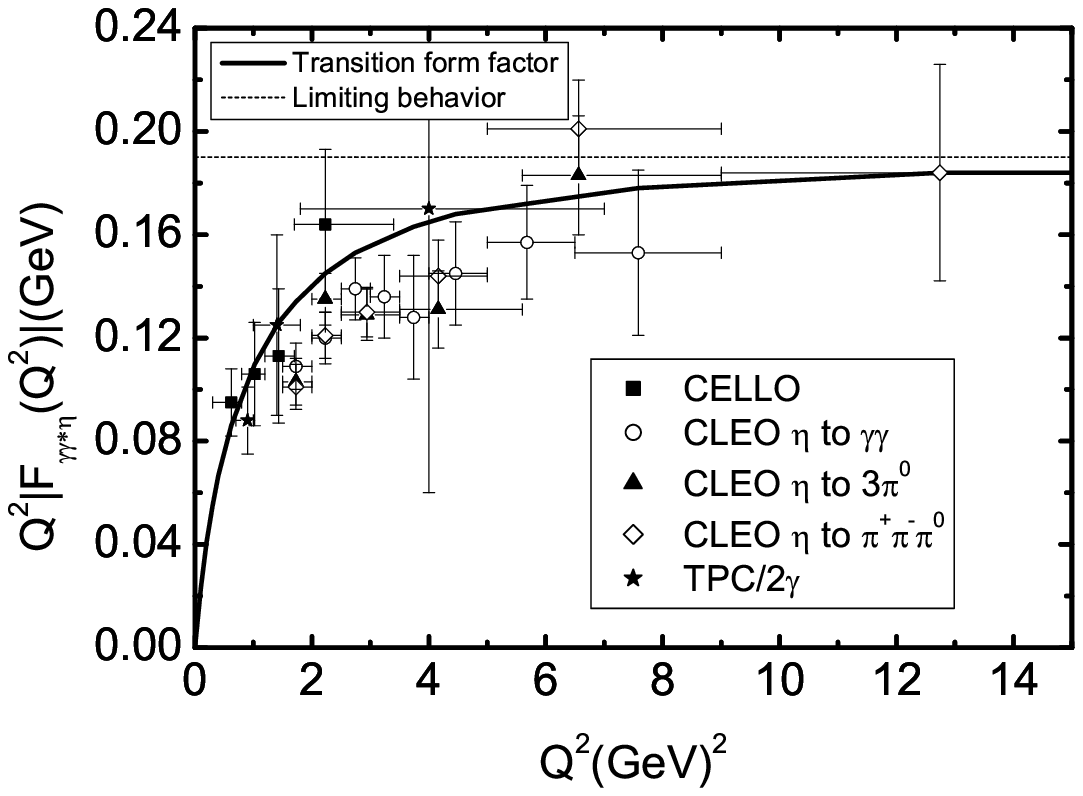}
\end{center}
\caption[*]{\baselineskip13pt Theoretical prediction of
$Q^{2}|F_{\protect\gamma \protect\gamma ^{\ast }\rightarrow
\protect\eta}(Q^{2})|$ compared with the experimental data. The
data for the transition form factor are taken from
Ref.~\protect\cite{CLEO98}, Ref.~\protect\cite{Beh91} and
Ref.~\protect\cite{TPC90}.} \label{eta2}
\end{figure}

\begin{figure}
\par
\begin{center}
\includegraphics{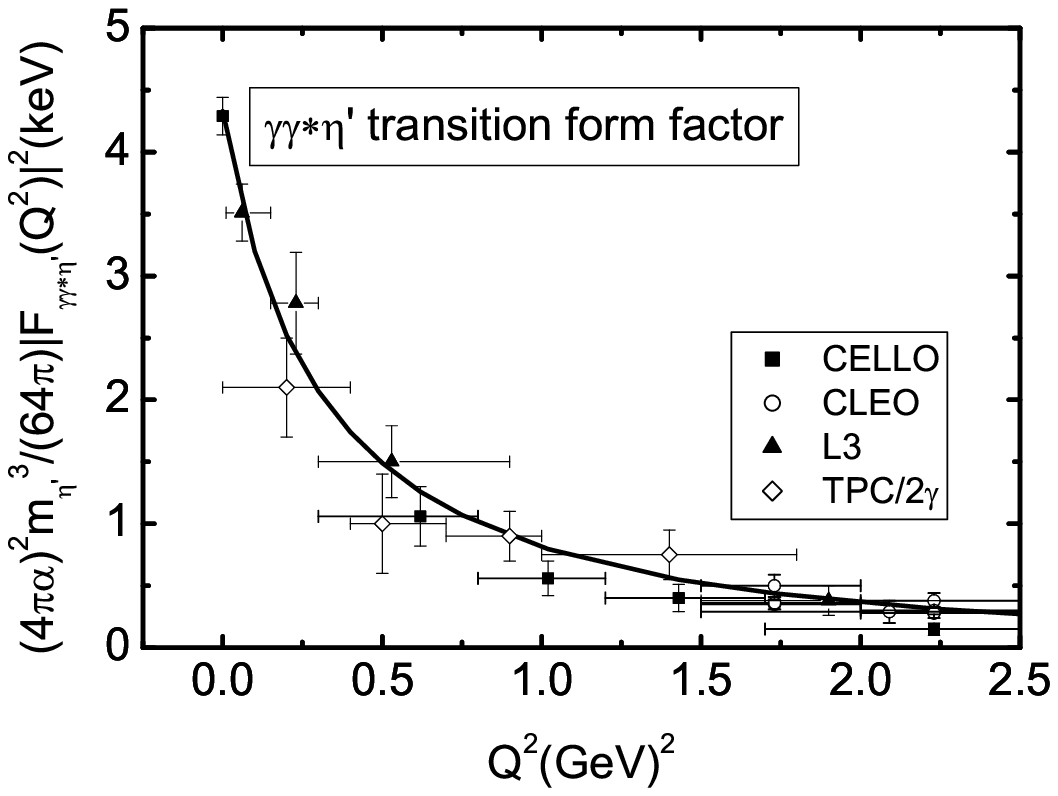}
\end{center}
\caption[*]{\baselineskip13pt Theoretical prediction  of $(4
\protect\pi \protect\alpha )^{2}\frac{m_{\protect\eta \prime
}^{3}}{64
\protect\pi }|F_{\protect\gamma \protect\gamma ^{\ast }\rightarrow \protect%
\eta \prime }(Q^{2})|^{2}$ compared with the experimental data in
the low energy scale. The data for the transition form factor are
taken from Ref.~ \protect\cite{CLEO98}, Ref.~\protect\cite{Beh91},
Ref.~\protect\cite{TPC90} and Ref.~\protect\cite{L398}.}
\label{etaprime1}
\end{figure}

\begin{figure}
\par
\begin{center}
\includegraphics{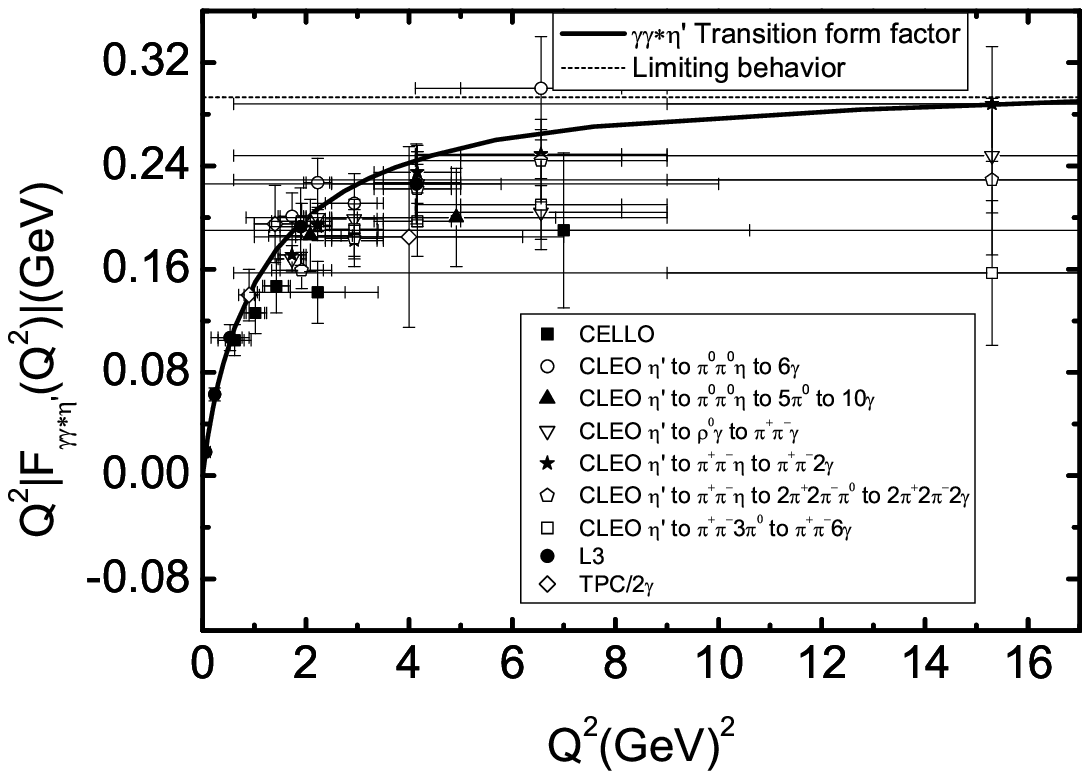}
\end{center}
\caption[*]{\baselineskip13pt Theoretical prediction of $%
Q^{2}|F_{\protect\gamma \protect\gamma ^{\ast }\rightarrow
\protect\eta^{\prime }}(Q^{2})|$ compared with the experimental
data. The data for the transition form factor are taken from
Ref.~\protect\cite{CLEO98}, Ref.~\protect\cite{Beh91},
Ref.~\protect\cite{TPC90} and Ref.~\protect\cite{L398}.}
\label{etaprime5}
\end{figure}

The results are in good agreement with the experimental data which
we have listed above. Moreover, it is interesting to notice that
the masses of the light-flavor quarks (the up quarks and down
quarks) from the above constrains are just in the correct range
(\textit{e.g.}, $200\sim 300~\mbox{MeV}$) of the constituent quark
masses from more general considerations. Naturally, the transition
form factor results emerging from this assumption are in quite
good agreement with the experimental data.

Fig.~2 indicates that the theoretical values of the photon-pion
$(\gamma \gamma ^{\ast }\rightarrow \pi )$ transition form factors
in the case of low $Q^{2}$ fit the experimental data well. One may
consider this work as a light-cone version of relativistic quark
model \cite{Ma93,Xiao2002}, which should be valid in the
low-energy scale about $Q^{2}\leq 2~\mbox{GeV}^{2}$. However, it
is also physically in accordance with the light-cone perturbative
QCD approach \cite{Lep80,Cao96}, which is applicable at the
high-energy scale of $Q^{2}>2~\mbox{GeV}^{2}$. The reason is that
the hard-gluon exchange between quark and antiquark of the meson,
which should be generally considered at high $Q^{2}$ for exclusive
processes, is not necessary to be incorporated in the light-cone
perturbative QCD approach for pion-photon transition form factor
\cite{Lep80,Cao96}. As a result, there is no wonder that our
predictions for the transition form factor at high $Q^{2}$ also
agree with the experimental data at high energy scale in Fig.~3.

Fig.~4 and Fig.~5 show that our predictions for the $\gamma ^{\ast
}\gamma \rightarrow \eta $ transition form factors agree with the
experimental data in the low and high energy scale, respectively.
In addition, the numerical results of $\gamma ^{\ast }\gamma
\rightarrow \eta ^{\prime }$ transition form factor also give good
fit of the experiments both in the low and moderately high energy
scale in Fig.~6 and Fig.~7. The prediction that we have made for
the low $Q^{2}~(0.001-0.5~\mbox{GeV}^{2})$ behaviors of the
photon-meson transition form factors of $\pi ^{0}$, $\eta $ and
$\eta ^{\prime }$ are measurable in
$e+A(\mbox{Nucleus})\rightarrow e+A+M$ process via Primakoff
effect at JLab and DESY.

Generally speaking, the medium to high $Q^2$ behavior of the
transition form factors should include leading-twist order
(so-called pQCD picture) and NLO corrections~\cite{Bra83,Kro2003},
but we only take the leading order into account in this
literature. However, we find that our results for the leading
order of the transition form factors fit the experimental data at
small $Q^{2}$ well and are also physically consistent with the
light-cone pQCD approach at large $Q^{2}$.

\section{Conclusion}

The light-cone formalism provides a convenient framework for the
relativistic description of hadrons in terms of quark and gluon
degrees of freedom, and for the application of perturbative QCD to
exclusive processes. With the minimal Fock-state expansions of the
pion and photon wave functions from the light-cone representation
of the spin structure of the pseudoscalar meson and photon
vertexes, we investigate the photon-meson transition form factors
by adopting the Drell-Yan-West assignment to get the light-cone
framework that works at both low $Q^2$ and high $Q^2$.
We employ the experimental values of the decay widths of $\pi $, $%
\eta $ and $\eta ^{\prime }$, the limiting behavior of
$\lim_{Q^{2}\rightarrow \infty }Q^{2}F_{\gamma ^{\ast }\gamma
\rightarrow M}(Q^{2})F_{\gamma ^{\ast }\gamma \rightarrow M}(0)$
$(M=\pi ,\eta _{8},\eta _{0})$, and the limiting behavior of
$Q^{2}F_{\gamma ^{\ast }\gamma \rightarrow M}(Q^{2})$ as the nine
constrains to fix those seven parameters in the $\pi $, $\eta
_{8}$, and $\eta _{0}$ wave functions. With the fixed $\pi $,
$\eta _{8}$, and $\eta _{0}$ wave functions, we find that our
numerical predictions for the photon-meson transition form factors
of light pseudoscalar mesons $\pi $, $\eta $, and $\eta ^{\prime
}$ agree with the experimental data at both low and high energy
scale, in a wide region comparing to previous studies.

\textbf{Acknowledgements} We acknowledge the helpful and
encouraging communications with Liping Gan. This work is partially
supported by National Natural Science Foundation of China under
Grant Numbers 10025523 and 90103007. It is also supported by
Hui-Chun Chin and Tsung-Dao Lee Chinese Undergraduate Research
Endowment (Chun-Tsung Endowment) at Peking University.


\begin{thebibliography}{99}
\bibitem{Lep80} G.P. Lepage and S.J. Brodsky, Phys.~Rev.~ {D 22},
 2157 (1980).

\bibitem{Cao96} F.-G. Cao, T. Huang, and B.-Q. Ma,~Phys.~Rev.  {D
53}, 6582
 (1996).

\bibitem{Xiao03} B.-W. Xiao and B.-Q. Ma, Phys. Rev. {D 68},
034020  (2003).

\bibitem{CLEO98} CLEO Collaboration, J. Gronberg \textit{et al.}, Phys. Rev.   {D 57}, 33 (1998).

\bibitem{Beh91} CELLO Collaboration, H.-J. Behrend \textit{et al.}, Z. Phys.
 {C 49}, 401 (1991).

\bibitem{Gan}
L. Gan, private communications, and proposal by PrimEx
Collaboration at JLab.

\bibitem{hem98}
HERMES Collaboration, K. Ackerstaff {\it et al.}, Nucl. Instr. and
Meth. {A 417}, 230 (1998).

\bibitem{Cao99} F.-G. Cao and A.I. Signal, Phys. Rev. {D 60}, 114012 (1999).

\bibitem{Feld99} T. Feldmann, P. Kroll, and B. Stech, Phys.\ Lett.\  {B}
 {449}, 339 (1999).


\bibitem{Chpt} J.L. Goity, A.M. Bernstein and B.R. Holstein, Phys. Rev. {D
66}, 076014 (2002).

\bibitem{nChpt} P. Maris and P.C. Tandy, Phys. Rev. {C65}, 045211
(2002); F.J. Llanes-Estrada and S.R. Cotanch, Nucl. Phys. {A 697},
303 (2002).



\bibitem{Cheng} Ta-Pei Cheng and Ling-Fong Li, ``Gauge theory of elementary
particle physics" (Claredon: Oxford, 1984).


\bibitem{Mixing} N. Beisert and B. Borasoy Eur. Phys. J. {A 11},
329 (2001); P. Herrera-Siklody, J.L. Latorre, P. Pascual and J.
Taron, Phys. Lett. {B 419}, 326 (1998).

Ll. Ametller, J. Bijnens, A. Bramon, and F. Cornet, Phys. Rev.
{\bf D 45}, 986 (1992).



\bibitem{Ani04} V.V. Anisovich, L.G. Dakhno, V.N. Markov, V.A. Nikonov, A.V.
Sarantsev, hep-ph/0410361; A.V. Anisovich, V.V. Anisovich, L.G.
Dakhno, V.A. Nikonov, A.V. Sarantsev. hep-ph/0406320.

\bibitem{Qiao00} S. Gieseke and C.F. Qiao, Phys. Rev. {D
61}, 074028 (2000).


\bibitem{Bro89} S.J. Brodsky, in \textit{Lectures on Lepton Nucleon
Scattering and Quantum Chromodynamics}, edited by A. Jaffe, D. Ruelle (Birkh$%
\bar{a}$user, Boston, 1982), p.255;

S.J. Brodsky and G.P. Lepage, in \textit{%
Perturbative Quantum Chromodynamics}, edited by A.H. Mueller,
(World Scientific, Singapore, 1989), p.93;

S.J. Brodsky, H-C. Pauli, and S.S. Pinsky, Phys. Rep.  {301},
 299 (1998).


\bibitem{BD80} S.J. Brodsky and S.D. Drell, Phys.\ Rev.\ D  {22}, 2236
(1980).

\bibitem{Bro2001} S.J. Brodsky, D.S. Hwang, B.-Q. Ma, and I. Schmidt, Nucl.
Phys.  {B 593}, 311 (2001).

\bibitem{BHL81} S.J. Brodsky, T. Huang, and G.P. Lepage, in \textit{\
Particles and Fields-2}, Proceedings of the Banff Summer
Institute, Banff, Alberta, 1981, edited by A.Z. Capri and A.N.
Kamal (Plenum, New York, 1983), p. 143.

\bibitem{Ma93} B.-Q. Ma, Z.~Phys. {A 345}, 321 (1993);

T. Huang, B.-Q. Ma, and Q.-X. Shen, Phys. Rev. D  {%
49}, 1490 (1994);

B.-Q. Ma and T. Huang, J. Phys. G: Nucl. Part. Phys.  {%
21}, 765 (1995);

F.-G. Cao, J. Cao, T. Huang, and B.-Q. Ma,~Phys.~Rev.  {D 55},
 7107 (1997).

\bibitem{Xiao2002}
B.-W. Xiao, X. Qian, and B.-Q. Ma, Eur. Phys. J.  {%
\ A 15}, 523 (2002).

\bibitem{Ma91} B.-Q.~Ma, J.Phys.G: Nucl. Part. Phys. {17}, L53 (1991);

B.-Q.~Ma and Q.-R.~Zhang, Z.~Phys. {C 58}, 479 (1993).

\bibitem{Ma96} B.-Q.~Ma, Phys.~Lett.~ {B 375}, 320 (1996);

B.-Q.~Ma and A.~Sch\"afer, Phys.~Lett.~ {B 378}, 307  (1996);

B.-Q.~Ma, I.~Schmidt, and J.~Soffer, Phys. Lett.  {B 441}, 461
(1998);

B.-Q.~Ma, I.~Schmidt, and J.-J. Yang, Eur. Phys. J.  {A 12}, 353
(2001).

\bibitem{MS98} B.-Q.~Ma and I.~Schmidt, Phys. Rev. {D 58},
 096008 (1998).

\bibitem{Don2001}
A. Donnachie, J. Gravelis, and G. Shaw, Phys. Rev.  {D 63}, 114013
(2001).


\bibitem{Hwang2001} C.-W. Hwang, Eur. Phys. J.  {C 19}, 105 (2001);

H.M. Choi and C.R. Ji, Nucl. Phys.  {A 618}, 291 (1997).

\bibitem{DYW} S.D.~Drell and T.-M. Yan, Phys. Rev. Lett. {24},
 181 (1970);

G. West, Phys. Rev. Lett. {24}, 1206 (1970).


\bibitem{Feld98} T. Feldmann and P. Kroll, Eur. Phys. J. {C 5}, 327 (1998).

\bibitem{Cao98} J. Cao, F.-G. Cao, T. Huang, and B.-Q. Ma,~Phys.~Rev.
 {D 58}, 113006 (1998).





\bibitem{Rad97}
A.V. Radyushkin and R. Ruskov, Phys.\ Lett.\  {B} {374}, 173
(1996).

\bibitem{LitTran}
See, \textit{e.g.}, R. Jakob, P. Kroll, and M. Raulfs, J. Phys.G:
Nucl. Part. Phys. {22}, 45 (1996);

V.V. Anisovich, D.I. Melikhov, and V.A. Nikonov, Phys. Rev. D
 {55}, 2918 (1997);

I.V. Musatov and A.V. Radyushkin, Phys.\ Rev.\ D { 56}, 2713
(1997).


\bibitem{PDB} Particle Data Group, K. Hagiwara {\it et al.}, Phys. Rev.  {D 66}, 010001 (2002).



\bibitem{Bro81} S.J. Brodsky and G.P. Lepage, Phys. Rev.  {D 24},
 1808 (1981).


\bibitem{decay} C. Caso et al., Eur. Phys. J.  {C 3}, 1 (1998).

\bibitem{TPC90} TPC/Two-Gamma Collaboration, H. Aihara \textit{et al.}, Phys. Rev. Lett. {64}, 172 (1990).


\bibitem{L398} L3 Collaboration, M. Acciarri \textit{et al.}, Phys. Lett. {B 418}, 399 (1998).


\bibitem{Bra83} E. Braaten, Phys. Rev.  {D 28}, 524
(1983).

\bibitem{Kro2003} P. Kroll and K. Passek-Kumericki, Phys. Rev. {D
67}, 054017 (2003).




\nonfrenchspacing
\end{thebibliography}
\end{document}